\documentclass[prd,showpacs,showkeys,preprint]{revtex4}

\usepackage{amsmath}
\usepackage{amssymb}
\usepackage{yfonts}[1988/10/03]
\usepackage{graphicx}
\newcommand {\beq}{\begin{equation}}
\newcommand {\eeq}{\end{equation}}
\newcommand {\beqa}{\begin{eqnarray}}
\newcommand {\eeqa}{\end{eqnarray}}

\begin{document}
\title{Lorentz violation and red shift of gravitational waves\\ in brane-worlds } 

\author{F. Ahmadi}
\email{fahmadi@srttu.edu}
\affiliation{ Department of Physics, Shahid
Rajaee Teacher Training University, Lavizan, Tehran 16788, Iran}
 \author{J. Khodagholizadeh} 
\email{j.gholizadeh@modares.ac.ir}
\affiliation{Departments of physics, Tarbiat Modares University, Tehran,14115-398, Iran.}
 \author{H. R. Sepangi}
\email{hr-sepangi@sbu.ac.ir} 
\affiliation{ Department of Physics, Shahid Beheshti University, G.
C., Evin, Tehran 19839, Iran.}

\begin{abstract}
In this paper we study the speed of gravitational waves in a  brane world
scenario and show that if the extra dimension is space-like, the
speed of the propagation of such waves is greater in the bulk than
that on the brane. Therefore, the $4D$ Lorentz invariance is
broken in the gravitational sector. A comparison is also made
between the red shift of such waves and those of the electromagnetic
waves on the brane. Such a comparison is essential for extracting
the signature of the extra dimension and thus clarifying the
question of maximal velocity of gravitational waves in the bulk.

\end{abstract}
\maketitle

\section{Introduction}

Considering the overall grounds we may expect that the theory of general relativity and quantum theory can be unified in the form of the standard model of particle physics at the Planck energy scale. However, this unification requires quantizing gravity, which results in very fundamental difficulties. One of these difficulties is related to the energy of the fundamental vacuum state and the other concerns  the status of Lorentz invariance. The fuzzy nature of space-time in quantum gravity may lead to violations of this fundamental symmetry. During the last two decades theoretical studies and experimental observation of Lorentz invariance violation have received a lot of attention \citep{Jac06, Mat05, Lib09}. One possible consequences of Lorentz invariance violation is energy dependent photon propagation velocities. The energy dependence can be constrained by recording the arrival times of photons of different energies emitted by distant objects at approximately the same time \citep{Amelino, Aharonian, Abdo}. One feature of Lorentz invariance violation to be considered is that the speed of light differs from that in special relativity. According to  gravity theories with Lorentz violation, the speed of graviton or the speed of gravitational wave differ from that in general relativity [see e.g.\citep{Pfeifer}]. Studying the speed of gravitational wave in a Lorentz violating gravity theory will provide different perspectives on quantum gravitational phenomena.

Of promising theories dealing with gravity  in recent years are the brane-world scenarios, offering a phenomenological way to test some of the novel predictions and corrections to general relativity that are implied by the M-theory. In such models, it is usually assumed that $c$ is a universal constant. Alternatively, approaches where the speed of gravitational waves can be different from $c$ in a brane-world context have also been considered, see\citep{Dvali,Deff,Damour}. It should be emphasized that the assumption that the maximal velocity in the bulk coincides with the speed of light on the brane should not be taken for granted. In this regard, theories with two metric tensors have been suggested with the associated two sets of ``null cones", in the bulk and on the brane \citep{merab}. This is the manifestation of violation of the bulk Lorentz invariance by the brane solution. In some brane-world scenarios, the space-time globally violates $4D$ Lorentz invariance, which results in apparent violations of Lorentz invariance from the brane observer's point of view due to bulk gravity effects. These effects are restricted to the gravity sector of the effective theory while the well measured Lorentz invariance of particle physics remains unaffected in these scenarios \citep{csaba,burgess, AJS}.

In this paper, we focus attention on the Einstein field equations on the brane obtained through the well known SMS procedure \citep{SMS} and address the question of the speed of propagation of gravitational waves in the bulk as well as on the brane. We find a relation between the maximum velocity in the bulk and the speed of light on the brane. Next, we compare the red shift experienced by gravitational waves traveling in the bulk with that of the electromagnetic waves on the brane and show that they are different. We find the signature of the extra dimension which would provide a possible detection mechanism in gravitational wave experiments, thus clarifying the maximal velocity in the bulk. If it turns out that the extra dimension is space-like, then we expect the Lorentz violation effects manifest themselves in gravitational wave experiments.
\section{Field equation}

In the usual  brane-world scenarios the space-time is identified
with a singular hypersurface (or 3-brane) embedded in a
five-dimensional bulk. Suppose now that the background manifold
$\bar{v_{4}}$ is isometrically embedded in a pseudo-Riemannian
manifold $v_{5}$ by the map ${\cal Y}:\bar{v_{4}}\rightarrow v_{5}$
such that
\begin{equation}
{\cal Y}^{A}_{,\mu}{\cal
Y}^{B}_{,\nu}g_{AB}=\bar{g_{\mu\nu}},\hspace{.3cm}{\cal
Y}^{A}_{,\mu}N^{B}g_{AB}=0,\hspace{.3cm} N^{A}N^{B}g_{AB}=\epsilon,
\label{eq1c}
\end{equation}
where $g_{AB}(\bar{g}_{\mu\nu})$ is the metric of the bulk (brane)
space $v_{5}(\bar{v}_{4})$ in arbitrary coordinate, ${{\cal
Y}^{A}}({{\cal X}^{\mu}})$ is the basis of the bulk (brane) and
$N^{A}$ is normal unite vector, orthogonal to the brane. Since
$N^{A}$ is a vector field along the extra dimension, we can
introduce $N^{A}$ as follows
\begin{equation}
N^{A}=\frac{\delta^{A}_{5}}{\phi}, \hspace{1.5cm}
N_{A}=(0,0,0,0,\epsilon\phi),\label{eq2}
\end{equation}
where $\phi$ is a scalar field. The perturbation  of $\bar{v}_{4}$
with respect to a small positive parameter $y$ along the normal
unit vector  $N^{A}$ is given by
\begin{equation}
{\cal Z}^{A}(x^{\alpha},y)={\cal Y}^{A}+y \phi (x^{\alpha},y)
N^{A}.\label{eq2c}
\end{equation}
The integrability conditions for the perturbed geometry are the
Gauss and Codazzi equations. The perturbation (\ref{eq2c}) induces a
perturbation on the metric $g_{\mu\nu}$ which can be written as
\begin{equation}
g_{\mu\nu}=\bar{g}_{\mu\nu}+{\cal
X}_{\mu\nu}(x^{\alpha},y).\label{eq3c}
\end{equation}
In particular, the linear perturbation obtained from the expansion
in $y$ is
\begin{equation}
g_{\mu\nu}=\bar{g}_{\mu\nu}+y
\gamma_{\mu\nu}(x^{\alpha}).\label{eq4c}
\end{equation}
To find the perturbed metric, we follow the same definitions as in
the geometry of surfaces. Consider the embedding equations of the
perturbed geometry written in the particular Gaussian frame
defined by the embedded geometry and the normal unit vector
\begin{equation}
{\cal Z}^{A}_{,\mu}{\cal
Z}^{B}_{,\nu}g_{AB}=g_{\mu\nu},\hspace{.3cm}{\cal
Z}^{A}_{,\mu}N^{B}g_{AB}=0,\hspace{.3cm}
N^{A}N^{B}g_{AB}=\epsilon. \label{eq5c}
\end{equation}
Substituting equation (\ref{eq2c}) in (\ref{eq5c}), we may express
the perturbed metric in the Gaussian frame defined by the embedding
as
\begin{equation}
g_{\mu\nu}=\bar{g}_{\mu\nu}+2 y \phi(x^{\alpha},y)
\bar{K}_{\mu\nu}+y^{2} \phi^{2}(x^{\alpha},y)
\bar{g}^{\alpha\beta}\bar{K}_{\mu\alpha}\bar{K}_{\nu\beta},\label{eq6c}
\end{equation}
where $\bar{K}_{\mu\nu}$ is the extrinsic curvature of the original
brane and the metric of our space-time is obtained at $y=0$
 $(g_{\mu\nu}=\bar{g}_{\mu\nu})$. Using equations
(\ref{eq2}), (\ref{eq5c}) and (\ref{eq6c}), the metric of the bulk
is written as
\begin{equation}
dS^{2}=g_{\mu\nu}(x^{\alpha},y)dx^{\mu}dx^{\nu}+\epsilon
\phi^{2}(x^{\alpha},y)dy^{2},\label{eq11}
\end{equation}
where we have used the signature $(+ - - - \epsilon)$ everywhere.
The five-dimensional Einstein equations contain the first and second
derivatives of the metric with respect to the extra coordinate.
These can be expressed in terms of geometrical tensors in $4D$. The
first partial derivatives can be written in terms of the extrinsic
curvature
\begin{equation}
K_{\mu\nu}=\frac{1}{2}{\cal
L}_{N}g_{\mu\nu}=\frac{1}{2\phi}\frac{\partial
g_{\mu\nu}}{\partial y}, \hspace{1cm}K_{A5}=0.\label{eq13}
\end{equation}
The second derivatives can be expressed in terms of the projection
$^{(5)}C_{\mu5\nu5}$ of the bulk Weyl tensor to 5$D$
\begin{eqnarray}
^{(5)}C_{ABCD} &=& ^{(5)\!\!\!}R_{ABCD}-\frac{2}{3}\left(^{(5)\!\!}R_{A[C}g_{D]B}-
^{(5)\!\!\!}R_{B[C}g_{D]A}\right)\ \nonumber\\ &+& \frac{1}{6}\left(^{(5)\!\!}R
g_{A[C}g_{D]B}\right).\label{eq14}
\end{eqnarray}
 In the absence of off-diagonal terms $(g_{5\mu}=0)$ the dimensional
reduction of the five-dimensional equations is particularly simple
\citep{ponce,JPL}. Thus, the field equations
can be split up into three parts
\begin{eqnarray}
^{(4)}G_{\mu\nu}&=&\frac{2}{3}k_{5}^{2}\left[^{(5)}T_{\mu\nu}+\left(^{(5)}T^{5}_{5}-\frac{1}{4}
(^{(5)}T)\right)g_{\mu\nu}\right]\nonumber\\
&-&\epsilon\left(K_{\mu\alpha}K^{\alpha}_{\nu}-K
K_{\mu\nu}\right)\nonumber\\&+&\frac{\epsilon}{2}g_{\mu\nu}\left(K_{\alpha\beta}K^{\alpha\beta}-K^{2}\right)-\epsilon
E_{\mu\nu}, \label{eq15}
\end{eqnarray}
\begin{equation}
\phi^{\mu}_{;\mu}=-\epsilon\frac{\partial K}{\partial y}-\phi
\left(\epsilon
K_{\alpha\beta}K^{\alpha\beta}+^{(5)\!\!}R^{5}_{5}\right),
\label{eq17}
\end{equation}
\begin{equation}
D_{\mu}(K^{\mu}_{\nu}-\delta^{\mu}_{\nu}K)=
k^{2}_{(5)}\frac{^{(5)}T_{5\nu}}{\phi}.\label{eq18}
\end{equation}
In the above expressions, $E_{\mu\nu}$  is the electric part of
the Weyl tensor and the covariant derivatives are calculated with
respect to $g_{\mu\nu}$, {\it i.e.}, $D g_{\mu\nu}=0$. With the brane-world scenario in mind,
for deriving the Einstein field equations on
the brane it is assumed that the
five-dimensional energy-momentum tensor has the form
\begin{equation}
^{(5)}T_{AB}=\Lambda_{5}g_{AB}+ ^{(5)}T^{(brane)}_{AB},\label{eq19}
\end{equation}
where $\Lambda_{5}$ is the cosmological constant in the bulk and $
^{(5)}T^{(brane)}_{AB}$ is the energy-momentum tensor of the matter on the brane with
$^{(5)}T^{(brane)}_{AB} N^{A}=0$ and
\begin{equation}
^{(5)}T^{(brane)}_{AB}=\delta^{\mu} _{A}  \delta^{\nu} _{B} \tau_{\mu\nu} \frac{\delta(y)}{\phi}.\label{eq20}
\end{equation}
In the spirit of the brane world scenario, we assume $ Z_{2}$
symmetry about the brane, considered to be a hypersurface $\Sigma$
at $y=0$. Using $Z_{2}$ symmetry, the Israel's junction conditions
are written as
\begin{equation}
K_{\mu\nu}|_{\Sigma^{+}}=-K_{\mu\nu}|_{\Sigma^{-}}=-\frac{\epsilon
k^{2}_{5}}{2}\left[\tau_{\mu\nu}-\frac{1}{3}g_{\mu\nu}\tau\right].
\label{eq35}
\end{equation}
Then, from equation (\ref{eq18}) it follow that
\begin{equation}
(\tau^{\mu}_{\nu})_{;\mu}=0,
\label{eq36}
\end{equation}
showing that the energy-momentum tensor $ \tau_{\mu\nu}$ is
conserved on the brane and represents the total vacuum plus matter
energy-momentum. It is usually separated in two parts,
\begin{equation}
\tau_{\mu\nu}=\sigma g_{\mu\nu}+T_{\mu\nu}, \label{eq38}
\end{equation}
where $ \sigma$ is the tension of the brane in $ 5D $, which is
interpreted as the vacuum energy of the brane world and $T_{\mu\nu}$
represents the energy-momentum tensor of ordinary matter in $4D$.
Using equations (\ref{eq35}) and (\ref{eq38}), we obtain the
Einstein field equations with an effective energy-momentum tensor in
$4D$ as
\begin{eqnarray}
^{(4)}G_{\mu\nu}&=&\Lambda_{4}g_{\mu\nu}+8 \pi G
T_{\mu\nu}-\epsilon k^{4}_{5}\Pi_{\mu\nu}-\epsilon
E_{\mu\nu} , \label{eq40}
\end{eqnarray}
where
\begin{eqnarray}
\Lambda_{4}&=&\frac{1}{2}k^{2}_{5}\left(\Lambda_{5}-\epsilon\frac{k^{2}_{5}\sigma^{2}}{6}\right),\hspace{0.5cm}
8 \pi G=-\epsilon\frac{k_{5}^{4}\sigma}{6},\ \nonumber\\
\\\Pi_{\mu\nu}&=&\frac{1}{4}
T_{\mu\alpha}T^{\alpha}_{\nu}-\frac{1}{12}T
T_{\mu\nu}-\frac{1}{8}g_{\mu\nu}T_{\alpha\beta}T^{\alpha\beta}+\frac{1}{24}g_{\mu\nu}T^{2}.
\nonumber\label{eq41}
\end{eqnarray}

All these $4D$ quantities have to be evaluated in the limit $
y\rightarrow 0^{+}$. They give a working definition of the
fundamental quantities $ \Lambda_{4}$ and $G$ and contain
higher-dimensional modifications to general relativity.

Also, to obtain the equations of gravitational waves in the bulk, first we assume
that the bulk space is empty. Using equation (\ref{eq6c}), the metric of the perturbed brane can
be written as fallow
\begin{equation}
g_{\mu\nu}=\eta_{\mu\nu}+\xi K_{\mu\nu},  \label{eq43a}
\end{equation}
where $\xi$ is a small parameter. By using the Einstien gauge, the equations of
gravitational waves become
\begin{equation}
 \Box K_{\mu\nu}= 0, \label{eq43b}
\end{equation}
since $ K_{\mu\nu} $ is related to the energy-momentum tensor on brane $(\tau_{\mu\nu})$ by
junction conditions, so we can conclude these waves are generated by ordinary matter on the brane.

\section{The bulk gravity effects and Lorentz violation }
In this section, we want to obtain a relation between the maximal velocity of propagation in
bulk and on the brane. Let us start by assuming a perfect fluid configuration on the brane.
The energy-momentum tensor is therefore written as
\begin{equation}
T_{\mu\nu}=(\rho+p)u_{\mu}u_{\nu}-pg_{\mu\nu}, \label{eq44}
\end{equation}
where $\textbf{u}$, $\rho$ and $p$ are the unit velocity, energy
density and pressure of the matter fluid respectively. We will
also assume a linear isothermal equation of state for the fluid
\begin{equation}
p=\gamma\rho,\hspace{10mm} 0\leq\gamma\leq 1. \label{eq45}
\end{equation}
The weak energy condition \citep{HE} imposes the restriction
$\rho\geq0$. In this paper we deal with non-tilted homogeneous
cosmological models on the brane, {\it i.e.} we are assuming that
the fluid velocity is orthogonal to the hypersurfaces  of
homogeneity. In standard cosmological models, we can also consider
$\phi(x^{\alpha}, y)=\phi(t)>0$ \citep{ponce}.

Next, we take the metric for our $4D$ universe as
\begin{equation}
\bar{g}_{\mu\nu}=\mbox{diag}(c_{b}^{2},-a(t)^{2}\Upsilon_{ij}),\label{eq46}
\end{equation}
with coordinates $(t,x^{i})$ and the 3-metric $\Upsilon_{ij}$ on
the spatial slices of constant time. Now, using the Israel's
junction condition, we have
\begin{eqnarray}
&\bar{K}_{00}&=\frac{\epsilon k_{5}^{2}\bar{g}_{00}}{6}\left[\sigma-(2+3\gamma)\rho\right],
\ \nonumber\\
\\
&\bar{K_{ii}}&=\frac{\epsilon k_{5}^{2}\bar{g}_{ii}}{3}\left[\sigma+\rho)\right].
\nonumber\label{eq48}
\end{eqnarray}
Substituting the above equations in equation (\ref{eq6c}), the
different $4D$ sections of the bulk in the vicinity of the original
brane will have the metric
\begin{equation}
g_{\mu\nu}=\Omega ^{2}\mbox{diag} (D c_{b}^{2},
-a(t)^{2}\Upsilon_{ij}),\label{eq47}
\end{equation}
where
\begin{eqnarray}
&\Omega^{2}&=\left[1+ \frac{\epsilon k_{5}^{2}}{6}y\phi(\sigma+\rho)\right]^{2}  ,
\nonumber\\ &D&=\left[\frac{6+\epsilon k_{5}^{2}y\phi(\sigma-(2+3\gamma)\rho)}{6+\epsilon k_{5}^{2}y\phi(\sigma+\rho) }\right].
\label{eq49}
\end{eqnarray}
From (\ref{eq46}), we see that the constant $c_{b}$ represents the
speed of light on the original brane, whereas from (\ref{eq47}) the
speed of the propagation of gravitational waves on the $4D$ section of the bulk is $D
c_{b}^{2}$. Now, in a brane world scenario where our universe is
identified with a singular hypersurface embedded in an $ AdS_{5}$
bulk, the extra dimension has to be space-like \citep{ponce}.
Therefore, $D$ is always greater than unity $(D>1)$. This leads to
apparent violations of Lorentz invariance from the brane observer's
point of view due to bulk gravity effects, since the maximal
velocity in the bulk becomes more than the speed of light on the
brane. It is worth noting that if the energy density of the matter
fluid on the brane is zero, we obtain $D=1$, implying that the
maximal velocity in bulk will be the speed of light on the brane.
On the other hand, this issue may lead to a modified dispersion relation
for propagating modes of gravitational waves. Thus, we should
construct a parameterized dispersion relation that can reproduce a range of
known Lorentz violation predictions. Also, we can investigate its effects on
how these waves propagate. In this regard, the reader is referred to \citep{sef, mir11}.

An interesting analogy exists between the behavior of gravitational
waves propagating into the bulk from the brane and the
electromagnetic waves crossing one medium into another with
different indexes of refraction. This is a reflection of Fermat's
principle where the greater speed achieved by gravitational waves in
the bulk is taken advantage of when such waves bend slightly into
the bulk with the result that they arrive earlier than
electromagnetic waves do, the latter being stuck to the brane.
Therefore, gravitational waves traveling faster than light would be
a possibility. These faster than light signals, however, do not
violate causality since the apparent violation of causality from the
brane observer's point of view is due to the fact that the region of
causal contact is actually bigger than what one would naively expect from
the ordinary propagation of light in an expanding universe, with no
closed timelike curves in the $5D$ spacetime that would make the
theory inconsistent \citep{CCEJ}.

\section{The red shift of gravitational waves }

As the present day observations of distant objects involve red
shifted spectra, knowing the behavior of the red shift of different
waves is necessary for analyzing data. In this section, we compare
the red shift of gravitational waves in the bulk with that of the
electromagnetic waves on brane. In doing so we consider the usual
FLRW line element on the brane with $a(t)$ as scale factor.
Therefore, the red shift of electromagnetic waves on the brane is
obtained by the usual formula
\begin{equation}
1+ z=\frac{\lambda}{\lambda_{0}}=\frac{a(t)}{a(t_{0})},\label{eq50}
\end{equation}
where $\lambda$ and $\lambda_{0}$ are the detected and  emitted
wavelengths, respectively. On the other hand, using equations
(\ref{eq47}), (\ref{eq49}), the red shift of gravitational waves on
brane is given by
\begin{equation}
1+ z=\frac{\lambda}{\lambda_{0}}= \sqrt{\frac{D(t_{0})}{D(t)}}\frac{a(t)}{a(t_{0})},\label{eq51}
\end{equation}
Now, the ratio of the red shift of the gravitational waves to that
of the electromagnetic waves is $\sqrt{\frac{D(t_{0})}{D(t)}}$ and
thus depends on the function $D(t)$. To study the variation
of the ratio with time, it is necessary to know the time variation
of $\rho$ and $\sigma $. From (\ref{eq36}) and (\ref{eq38}) it
follows that
\begin{equation}
\sigma_{,\nu}+T^{\mu}_{\nu:\mu}=0.\label{eq52}
\end{equation}
For the perfect fluid (\ref{eq44}), this is equivalent to
\begin{equation}
\dot{\rho}+(\rho+p)\Theta=-\dot{\sigma},\label{eq53}
\end{equation}
\begin{equation}
(\rho+p)a_{\nu}+(\gamma-1)\rho_{,\lambda}h^{\lambda}_{\nu}=\sigma_{,\lambda}h^{\lambda}_{\nu},\label{eq54}
\end{equation}
where $\Theta=u^{\mu}_{;\mu}$ is the expansion,
$a_{\nu}=u_{\nu;\lambda}u^{\lambda}$ is the acceleration and
$h_{\mu\nu}=u_{\mu}u_{\nu}-g_{\mu\nu}$ is the projector onto the
spatial surface orthogonal to $u_{\mu}$. In homogeneous
cosmological models, equation (\ref{eq54}) becomes redundant and
only equation (\ref{eq53}) is relevant. For the case of the constant
vacuum energy $\sigma$ and equation of state $p=(\gamma-1)\rho$, it
yields the familiar relation between the matter energy density and
expansion factor $a$
\begin{equation}
\rho=\rho_{0}\left(\frac{a_{0}}{a}\right)^{3\gamma}.\label{eq55}
\end{equation}
For the case where the vacuum energy is not constant, we need some
additional assumption. We know that the time variation of $G$ is
usually written as $(\frac{\dot{G}}{G})=g H$, where $g$ is a
dimensionless parameter. The present observational upper bound is
$|g|\leq 0.1$ \citep{ponce,JPL}. In what follows we assume that $g$
is constant. Since $G\sim\sigma$ and $H=\frac{\dot{a}}{a}$,  we have
$\sigma(a)=f_{0}a^{g}$, where $f_{0}$ is a constant of integration.
From equation (\ref{eq53}) we thus find
\begin{equation}
\dot{\rho}+3\gamma\rho\frac{\dot{a}}{a}=-f_{0} g a^{(g-1)}\dot{a}.\label{eq56}
\end{equation}
First, we consider the case where $\rho$ can be expressed in $a$,
similar to (\ref{eq55}), i.e. as a power of $a$. We therefore find $
g=\frac{-3\gamma E}{E+f_{0}}$, and
\begin{equation}
 \rho= E a^{-\left(\frac{3\gamma E}{E+f_{0}}\right)},\label{eq57}
\end{equation}
where $E$ is a positive constant. In order to simplify the notation we set $f_{0}=F_{0}E$ and
\begin{equation}
\frac{\gamma}{1+F_{0}}=\beta+1,\label{eq58}
\end{equation}
which gives
\begin{equation}
\beta=\frac{\gamma-F_{0}-1}{1+F_{0}}.\label{eq59}
\end{equation}
With this notation we now have
\begin{equation}
\rho=\frac{E}{a^{3(\beta+1)}},\hspace{1cm} \sigma=\frac{E(\gamma-1-\beta)}{(\beta+1) a^{3(\beta+1)}}.\label{eq60}
\end{equation}
Note that one may take $E=\rho_{0}a_{0}^{3(\beta+1)}$ and that
$F_{0}\neq 0$ $ (\beta\neq\gamma-1)$, otherwise $G=0$. We also have
\begin{equation}
\frac{\dot{G}}{G}=-3 (\beta+1) H.\label{eq61}
\end{equation}
Before going any further, we should be aware of observational bounds
on $\beta$. The lower bound comes from the obvious requirement
$\frac{d\rho}{da}<0$, while the upper bound comes from the
observation that $|g|\leq0.1$. Thus,
\begin{equation}
-1<\beta\leq-0.966,\hspace{1cm} \beta=-0.983\pm0.016.\label{eq62}
\end{equation}
The next step would be to find the function $\phi(t)$. In 
standard cosmological models, we can set
$\phi=\phi_{0}(\frac{a}{a_{0}})^{r}$. The physical condition
$\tau^{0}_{0}=(\sigma+\rho)>0 $ then puts a lower limit on $r$,
namely $r>1$ \citep{ponce}. Therefore, we find $D(t)$ as
\begin{eqnarray}
D(t)&=&\left[1-\frac{3\epsilon k_{5}^{2}y\phi_{0}\rho_{0}(1+\gamma)
\left(\frac{a}{a_{0}}\right)^{r-3(\beta+1)}}{6+\epsilon
k_{5}^{2}y\phi_{0}\rho_{0}\left(\frac{\gamma}{\beta+1}\right)
\left(\frac{a}{a_{0}}\right)^{r-3(\beta+1)}}\right]\ \nonumber\\
&\simeq&\left[1-\frac{\epsilon}{2}k_{5}^{2}y\phi_{0}\rho_{0}(1+\gamma)
\left(\frac{a}{a_{0}}\right)^{r-3(\beta+1)}\right],\label{eq63}
\end{eqnarray}
and
\begin{equation}
\frac{D(t_{0})}{D(t)}=\frac{1-\frac{\epsilon}{2}k_{5}^{2}y\phi_{0}\rho_{0}(1+\gamma)}
{1-\frac{\epsilon}{2}k_{5}^{2}y\phi_{0}\rho_{0}(1+\gamma)
\left(\frac{a}{a_{0}}\right)^{r-3(\beta+1)}}.\label{eq64}
\end{equation}
Now, if the extra dimension is space-like,
$\sqrt{\frac{D(t_{0})}{D(t)}}$ is less than unity and so the red
shift due to gravitational waves is smaller than that  of the
electromagnetic waves. On the other hand, if the extra dimension is
time-like, $\sqrt{\frac{D(t_{0})}{D(t)}}$ is greater than unity.
Then, the red shift of gravitational waves is greater than that of
the electromagnetic waves. 

Since the red shift of gravitational waves is
different from that of the electromagnetic waves, this issue may have effects on
the detection of gravitational waves. For instance,
through the process of finding a correlation between the small scale CMB polarization fluctuations
and the galaxy number density at a given red shift, one can determine the local
quadrupole moments of the CMB at that red shift. Then, considering these quadrupoles at different patches on the sky and at different red shifts, one can obtain a map of the quadrupole moments during the reionization
era \citep{Kam97}. A small part of this quadrupole pattern can be produced by the tensor modes of fluctuations, {\it i.e.},
gravitational waves. Using the correlation between galaxy distribution and
the CMB polarization anisotropies, we can constrain the strength of the primordial gravitational waves which are really
important to physicists, see also  \citep{Cooray, Ports, Bunn, Ali12}. Since the red shift of gravitational waves
is different from that of electromagnetic waves,
one expects the strength of primordial gravitational waves to be modified.

\section{Conclusions}
In this paper we have considered a brane-world scenario where the
Einstein field equations  on the brane were obtained using the usual
SMS formalism. Based on this scenario we then showed that within the
framework of the standard cosmological models, the red shift
associated with gravitational waves moving through the bulk is not
equal to the red shift of electromagnetic waves propagating on the
brane. Such a difference could be used, in the event of the
detection of gravitational waves, to find the signature of extra
dimension. This should clarify the question of the maximal velocity
in the bulk space. We also showed that if the extra dimension is
space-like, the $4D$ Lorentz invariance in the gravitational sector
is broken in the sense of having a propagation speed greater than
that of  light. Gauge fields will not feel these effects, but
gravitational waves are free to propagate into the bulk and they
will necessarily feel the effects of the variation of the speed of
light along the extra dimension.

\end{document}